\documentclass[conference]{IEEEtran}
\IEEEoverridecommandlockouts
\usepackage{cite}
\usepackage{amsmath,amssymb,amsfonts}
\usepackage{algorithmic}
\usepackage{graphicx}
\usepackage{textcomp}
\usepackage{xcolor}
\usepackage{pgfplotstable}
\usepgfplotslibrary{statistics}
\pgfplotsset{grid style={dotted,gray}}
\usepackage{tikz}
\usepackage{xcolor}
\pgfplotsset{compat=1.8}
\usepackage{pgfplots}
\usetikzlibrary{spy}
\def\BibTeX{{\rm B\kern-.05em{\sc i\kern-.025em b}\kern-.08em
    T\kern-.1667em\lower.7ex\hbox{E}\kern-.125emX}}
\begin{document}

\title{UAV-assisted Distributed Learning for Environmental Monitoring in Rural Environments\\

\thanks{This work has received funding from the Horizon 2020 research and innovation staff exchange grant agreement No 101086387, and from the  Science Fund of the Republic of Serbia, grant number 6707, REmote WAter quality monitoRing anD IntelliGence – REWARDING}
}

\author{\IEEEauthorblockN{Vukan Ninkovic,\IEEEauthorrefmark{1}\IEEEauthorrefmark{2} Dejan Vukobratovic,\IEEEauthorrefmark{1} {Dragisa Miskovic 
 \IEEEauthorrefmark{2}}
\IEEEauthorblockA{\IEEEauthorrefmark{1}University of Novi Sad, 
Novi Sad, Serbia}
\IEEEauthorblockA{\IEEEauthorrefmark{2}The Institute for Artificial Intelligence Research and Development of Serbia, Novi Sad, Serbia
}}}

\maketitle

\begin{abstract}
Distributed learning and inference algorithms have become indispensable for IoT systems, offering benefits such as workload alleviation, data privacy preservation, and reduced latency. This paper introduces an innovative approach that utilizes unmanned aerial vehicles (UAVs) as a coverage extension relay for IoT environmental monitoring in rural areas. Our method integrates a split learning (SL) strategy between edge devices, a UAV and a server to enhance adaptability and performance of inference mechanisms. By employing UAVs as a relay and by incorporating SL, we address connectivity and resource constraints for applications of learning in IoT in remote settings. Our system model accounts for diverse channel conditions to determine the most suitable transmission strategy for optimal system behaviour. Through simulation analysis, the proposed approach demonstrates its robustness and adaptability, even excelling under adverse channel conditions.  Integrating UAV relaying and the SL paradigm offers significant flexibility to the server, enabling adaptive strategies that consider various trade-offs beyond simply minimizing overall inference quality.
\end{abstract}


\section{Introduction}

The pervasive issue of pollution, stemming from various sources such as industrial activities, transportation emissions, and waste disposal, poses significant challenges to the environment, raising concerns about its impacts \cite{ullo_2020}. Ensuring health and hygiene is vital for both humanity's sustainability and a nation's progress, that is dependent on a clean, hazard-free environment. Therefore, monitoring these aspects is essential to promote a healthy life for citizens, especially in rural and underdeveloped environments. In recent years, the integration of IoT for environmental monitoring in rural areas has emerged, utilizing interconnected devices equipped with diverse sensors to gather real-time data on crucial environmental parameters like air quality, soil moisture, water quality, temperature, and humidity \cite{mois_2017, ullo_2020}. These devices employ wireless communication technologies such as Wi-Fi, cellular networks, LoRaWAN, or satellite connectivity to transmit data to centralized servers or cloud platforms for storage, analysis, and further processing \cite{mois_2017}.

Unmanned aerial vehicles (UAVs) have found widespread applications across various industries, governmental bodies, and commercial sectors, performing tasks ranging from telecommunications, rescue operations, to surveillance\cite{mozafir_2019, sabze_2022}. Notably, they have gained significant attention for their pivotal role in enabling end-to-end wireless communications, especially in providing connectivity to wireless (IoT) devices in remote or rural areas where traditional cellular coverage is scarce or absent \cite{sabze_2022}. Specifically, access points integrated onto UAVs are being proposed as a potential solution for anticipated data demand and congestion challenges in future wireless networks \cite{zheng_2016}. Unlike conventional static infrastructure, UAV networks offer the advantage of flexible deployment, enabling them to te coverage \cite{galkin_2018}.

In this work, we propose a novel approach for distributed learning in IoT environmental monitoring scenarios. Our method integrates the split learning (SL) paradigm with UAV relaying in an IoT network, enhancing data transmission rates and ensuring equitable distribution of computational tasks between edge devices, UAV, and server. Furthermore, our approach enhances system adaptability by empowering the server to determine the optimal transmission strategy based on current channel conditions and specific performance metrics such as latency, throughput, and energy efficiency. Numerical results demonstrate that by utilizing proposed distributed learning approach, IoT system shows great robustness to different channel conditions, simultaneously achieving high performance in terms of estimation accuracy.  

\section{Background}

\subsection{IoT Connectivity in Rural Areas}

According to \cite{yaacoub_2021}, IoT connectivity mostly depends on the level of development of the country. More precisely, in developed countries, rural areas are typically accessible via transportation networks, such as railroad networks, and power is supplied through the electricity grid. However, the challenge lies in mobile operators obtaining a satisfactory return on investment (ROI) for providing backhaul to these areas. Conversely, in developing countries, particularly impoverished areas, the challenge is to bridge the digital divide with developed nations. In rural areas, essential services like healthcare and education depend on connectivity, but inadequate transportation infrastructure isolates villages from major cities, while power generation often relies on local sources. Establishing backhaul in such areas, starting from scratch and facing limited revenue due to poverty, may require state subsidies to emphasize the necessity for cost-effective solutions.

 In the context of providing connectivity in rural areas, especially in developing countries, unmanned aerial vehicles (UAVs) equipped with communication equipment can play a significant role \cite{zhang_2019}. These drones can serve as flying base stations, establishing temporary or permanent connectivity in areas where traditional infrastructure deployment is impractical or cost-prohibitive. By flying over remote regions, drones can establish wireless links between users and the broader network infrastructure, effectively extending backhaul links to underserved communities  \cite{fouda_2018}. In regions with underdeveloped or non-existent transportation infrastructure, such as in impoverished areas of developing countries, drones offer a versatile and efficient means of providing backhaul links \cite{yaacoub_2021}. They can be swiftly deployed and are adaptable to changing conditions, making them valuable in emergency situations or areas with limited access to resources \cite{selim_2018}.

\subsection{Split Learning}
\label{SL}
Split learning (SL) \cite{gupta_2018, vep_2018} is a new distributed learning paradigm, which  divides a neural network $F$ (consisting of $L$ layers) into sequential layers across multiple participants, like an edge device and a server. In SL, the edge device shares its training dataset securely with the server, which oversees the training process and handling most computational tasks. This distributed approach accelerates convergence and reduces bandwidth constraints \cite{vep_2018}.

SL separates model training and inference processes. During training, data remains within individual edge devices to prevent raw information transmission across the network. The neural network can be represented as $F=(f_{\text{E}}, f_{\text{S}})$, where $f_{\text{E}}:\mathbb{R}^{N}\rightarrow\mathbb{R}^{M}$ and $f_{\text{S}}:\mathbb{R}^{M}\rightarrow\mathbb{R}^{1}$,  ($N$ and $M$ are dimensions of raw data and intermediate representation, respectively, with $M<N$). During activation, the edge device sub--network produces an intermediate representation of raw data $\boldsymbol{x}$ as $\boldsymbol{z}=f_{\text{E}}(\boldsymbol{x})$, sent to the server for prediction $\hat{y}=f_{\text{S}}(\boldsymbol{z})$ ($f_{\text{S}}$ is sub--network deployed at the server side). This training enables collaborative learning without compromising data privacy, achieved by iteratively exchanging model updates during backward passes between the server and edge device \cite{vep_2018}.
In the realm of split inference, split learning optimizes efficiency by employing pre--trained models distributed across multiple devices. Initial data processing occurs locally on edge devices, generating intermediate representations $\boldsymbol{z}$, which are then transmitted to a centralized server for aggregation and final inference \cite{vep_2018}.

In the context of rural IoT network which, except edge device and server, consist of relaying drone (UAV) that provides a backhaul link, neural network $F$ is divided into three main parts, i.e., $F=(f_{\text{E}}, f_{\text{D}}, f_{\text{S}})$, where $f_{\text{D}}:\mathbb{R}^{M}\rightarrow\mathbb{R}^{M}$  represents drone sub--network.

\subsection{Recurrent Neural Networks}
\begin{figure}[t]
 \centerline{\includegraphics[width=1\columnwidth, height=1.8in]{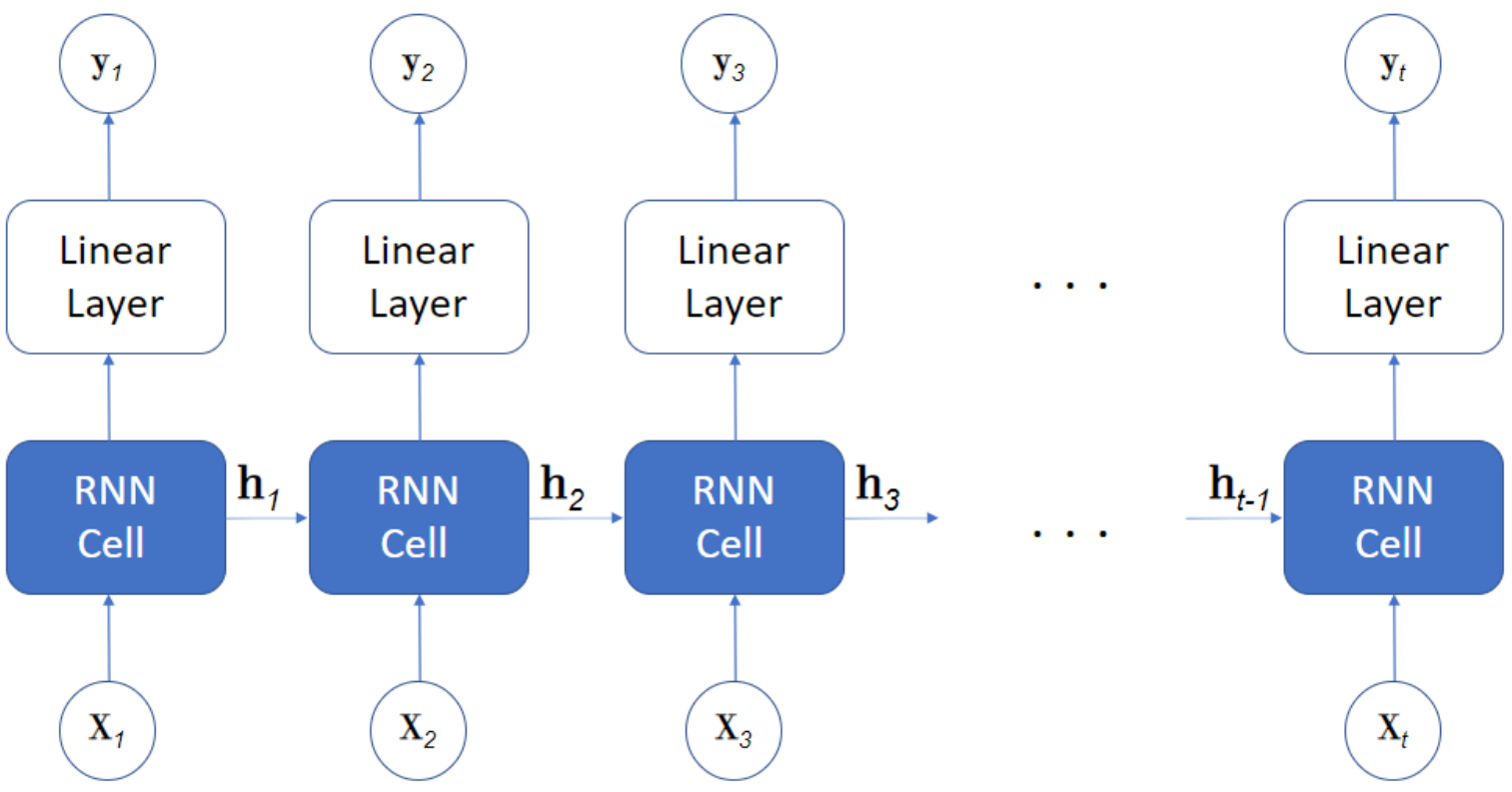}}
  \caption{The structure of Recurrent Neural Network.}
  \label{fig:RNN}
\end{figure}

RNNs, as sequence-based models, have the capability to discern temporal relationships between preceding and current states. Consequently, they represent an ideal solution for processing time series data \cite{husken_2002}. Fig. \ref{fig:RNN} presents a simple depiction of a single-layer RNN. In this illustration, the output from the previous time step, denoted as $t-1$, is incorporated into the input of the current time step, denoted as $t$, thus enabling the retention of past information. The computation outcome of a single RNN cell can be described by the following function:

\begin{equation}
    \boldsymbol{h}_{\text{t}} = \tanh(\boldsymbol{W}_{\text{ih}}\boldsymbol{x}_{\text{t}} + \boldsymbol{b}_{\text{ih}} + \boldsymbol{W}_{\text{hh}}\boldsymbol{h}_{\text{t-1}} + \boldsymbol{b}_{\text{hh}}),
\end{equation}
where $\tanh$ denotes the hyperbolic tangent function, $\boldsymbol{h}_t$ and $\boldsymbol{h}_{t-1}$ represent the hidden states at time steps $t$ and $t-1$, respectively, while $\boldsymbol{W}_{ih}$, $\boldsymbol{W}_{hh}$, $\boldsymbol{b}_{ih}$, and $\boldsymbol{b}_{hh}$ are the weights and biases requiring learning, with $\boldsymbol{x}_{t}$ denoting the input at time $t$.

Basic RNN cells encounter challenges in learning long-range dependencies primarily due to issues such as vanishing or exploding gradients. To address this limitation, Long Short-Term Memory (LSTM) cells were introduced, as proposed in \cite{b13}. These cells incorporate specialized units known as memory blocks within the recurrent hidden layer, thereby augmenting their ability to capture long-term dependencies. Each memory block constitutes a recurrently connected sub--network comprising functional components, namely memory cells and gates. The memory cells retain temporal states of the network, whereas the gates regulate the flow of information from the preceding cell state.

\subsubsection{Split Learning--Based RNNs}
Initially integrating the split learning/inference paradigm into LSTM neural networks faced challenges, prompting researchers to seek alternative methods. Some studies have advocated for using 1D-CNN instead of LSTMs to tackle these issues effectively \cite{abua_2020, Gao_2020, zhang_2024}. Recent research has introduced efficient approaches to integrate split learning into LSTM networks \cite{koda_2020, jiang_2022, abedi_2023}, embedding the split learning paradigm directly into LSTM architectures to overcome implementation obstacles with innovative strategies.

This paper builds upon the foundational work of \cite{jiang_2022}, which introduced the \textit{LSTMSPLIT} algorithm, splitting the LSTM neural network vertically, and requiring a minimum of two LSTM layers, with the input sequence stored at the edge device. Following the procedure outlined in Section \ref{SL}, the intermediate representation $\boldsymbol{z}$ is transmitted from the edge device's LSTM layer to the server's LSTM layer, while update gradients move in the opposite direction (as depicted on Fig. \ref{fig:sys}). In the second approach \cite{abedi_2023}, the authors proposed a method where one LSTM layer is distributed across multiple edge devices, partitioned into sub-networks trained individually on each device. This enables the handling of segments within multi-segment training sequences.
 Communication among edge devices and parameter sharing facilitate inference, aligning with the federated learning paradigm.

 \section{UAV--assisted Relaying in IoT}
 
 \subsection{System Model}
 \label{model}
\begin{figure}[t]
 \centerline{\includegraphics[width=0.8\columnwidth, height=1.8in]{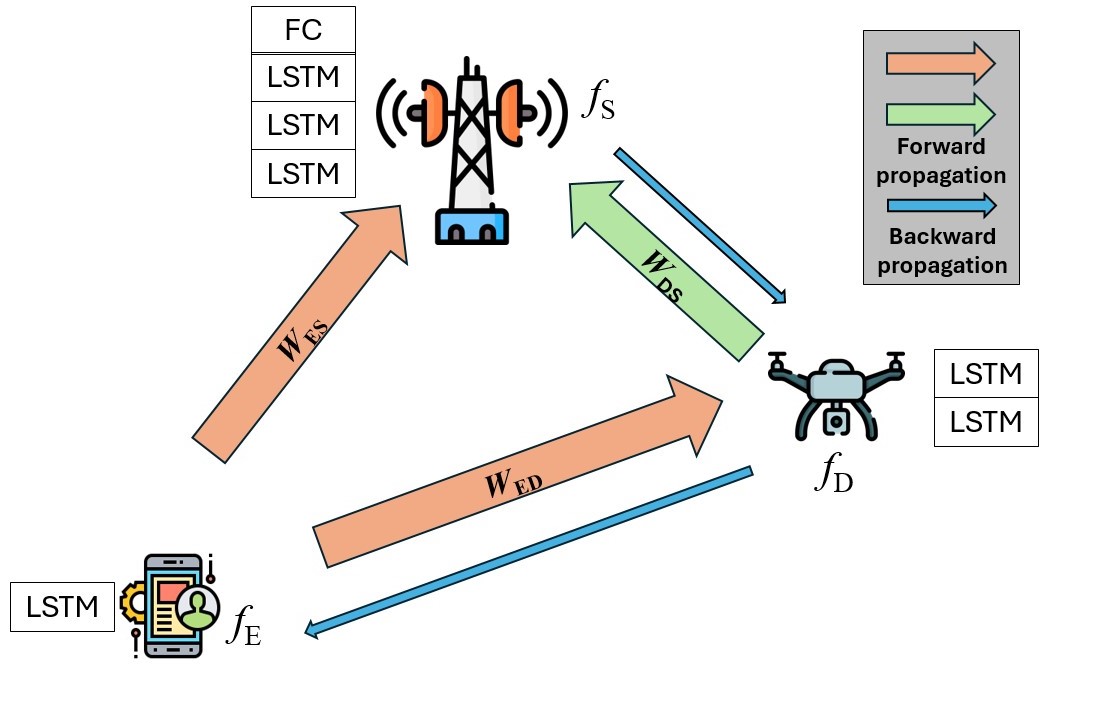}}
  \caption{SL--based fronthaul/backhaul communication with different channel conditions.}
  \label{fig:sys}
\end{figure}

We examine the conventional IoT system, consisting of an edge device, server, and UAV, where the drone serves as a wireless relay, effectively functioning as a base station with backhaul link \cite{fouda_2018} (see Fig. \ref{fig:sys}). Each of these devices possesses unique computational capabilities, i.e, $\mathcal{C}(f_{\text{E}})<\mathcal{C}(f_{\text{D}})<\mathcal{C}(f_{\text{S}})$ (where $\mathcal{C}(\cdot)$ is sub--network complexity). At the edge device, we gather raw data denoted by $\boldsymbol{x}$, which may comprise sensor measurements, and these data points are labeled with corresponding labels $y$. Subsequently, this raw data undergoes pre--processing by the edge device sub--network, resulting in the intermediate representation $\boldsymbol{z}=f_{\text{E}}(\boldsymbol{x})$.

In this study, we establish assumptions concerning various levels of autonomy, primarily concentrated on the server side, which assumes responsibility for coordinating communication and inference processes. Specifically, taking into account channel conditions and essential performance metrics such as error rate, latency, and communication overhead, the server determines whether direct communication with the edge device is warranted or if the intermediate representation should undergo further processing by a drone sub-network. Furthermore, we anticipate significant variations in channel conditions, potentially differing between the edge device and drone (red arrow, $\mathcal{W}_{\text{ED}}$ on Fig. \ref{fig:sys}) as well as between the edge device and server ( red arrow, $\mathcal{W}_{\text{ES}}$ on Fig. \ref{fig:sys}). Notably, we expect the channel between the drone and server (green arrow, $\mathcal{W}_{\text{DS}}$, on Fig. \ref{fig:sys}) to maintain consistently good quality throughout the entire network's lifespan. 

 Regarding the implemented strategy, the intermediate representation $\boldsymbol{z}$ encounters various channel conditions, resulting in its distorted version arriving at the server side. If direct communication occurs between the edge device and server, the intermediate representation at the server input can be defined as $\hat{\boldsymbol{z}}=\mathcal{W}_{\text{ES}}(\boldsymbol{z})$.
Conversely, if communication between the edge device and server traverses a drone backhaul link, and the drone sub--network is involved in overall processing and prediction, then $\hat{\boldsymbol{z}}=\mathcal{W}_{\text{DS}}(f_{\text{D}}(\mathcal{W}_{\text{ED}}(\boldsymbol{z})))$. On the server side, local decisions are made regarding desirable performance criteria, particularly latency constraints. For example, the server determines the optimal strategy, deciding whether the entire server sub--network will be included in the prediction process or only its output layer. More precisely, the estimation of $y$ can be defined as $\hat{y}_{full}=f_{\text{S}}(\hat{\boldsymbol{z}})$ or $\hat{y}_{FC}=\hat{f}_{\text{S}}(\hat{\boldsymbol{z}})$  if the server sub-network comprises all layers or only its output layer, respectively. 

According to obtained estimation, appropriate loss function, which consists of two different parameters is calculated at server side as: 
\begin{align}
\label{loss}
    \mathcal{L}(y, \hat{y}_{\text{full}}, \hat{y}_{\text{FC}})=MSE(y,\hat{y}_{\text{full}})+MSE(y,\hat{y}_{\text{FC}}),
\end{align}
where $MSE(y, \hat{y})=1/|\mathcal{D}_{\text{tr}}
|\sum_{\mathcal{D}_{\text{tr}}}(y-\hat{y})^2$ is mean--squared error (MSE), $\mathcal{D}$ is training dataset and $|\cdot|$ is its cardinality.

 In the backward propagation phase, gradients are determined with respect to the loss function and then conveyed from the server, passing through a drone, and finally directed towards the edge device, essentially reversing the neural network's direction (as illustrated by the blue arrows in Fig. \ref{fig:sys}). In such a case, all three sub--networks ($f_{\text{E}}$, $f_{\text{D}}$, $f_{\text{S}}$) are jointly optimized. The collaborative optimization typically involves fine-tuning the parameters of all three sub--networks using optimization algorithms like stochastic gradient descent (SGD) or its adaptations, such as Adam \cite{adam}.

Under the above--mentioned setup, the main goal here is to define the most suitable transmission strategy, regarding the channel conditions and desirable latency, for minimization overall system error, defined as MSE error from Eq. (\ref{loss}), between $y$ and $\hat{y}$ across all test examples.

 \subsection{Channel Model}
We consider relatively simple wireless communication link between edge device and server ($\mathcal{W}_{\text{ES}}$), edge device and drone ($\mathcal{W}_{\text{ED}}$) and drone and server ($\mathcal{W}_{\text{DS}}$). These links are modeled as conventional erasure channels, with an erasure probability denoted by $p$. This channel can be represented as a binary vector $\boldsymbol{q} \in \{0,1\}^M$, where $M$ is the length of the intermediate representation $\boldsymbol{z}$ (as discussed in Section \ref{SL}). Individual symbols from $\boldsymbol{z}$ are either erased or they arrive unchanged at the server side. Consequently, $\hat{\boldsymbol{z}}=\boldsymbol{z}\odot\boldsymbol{q}$, where $\odot$ represents element--wise multiplication \cite{itahara_2022}.

\section{Performance Evaluation}

\subsection{Training Setup}
To evaluate the proposed approach and assess the influence of different channel conditions on overall system performance, as well as the significance of the backhaul, we utilized a dataset perfectly suited to the environmental problem of interest. Specifically, we focus on monitoring pollution in the Danube river near Novi Sad. Our dataset comprises 3,264 instances, with 70\%\ utilized for training and the remaining 30\%\ used for testing purposes. Each instance represents a daily measurement from November 2013 to October 2022, encompassing eight different water quality parameters: temperature, pH value, electrical conductivity, dissolved oxygen, oxygen saturation, ammonium, and nitrite.

Based on the correlation matrix between all measured features, we have decided to predict dissolved oxygen using its last 20 measurements (over the previous 20 days) along with measurements of the other 7 parameters for the current day. More precisely, following data preprocessing and conversion to time series, each instance in the dataset comprises 27 features (including 20 previous dissolved oxygen measurements and 7 other parameters) and one label (representing dissolved oxygen for the current day). Additionally, considering that different parameters are measured on varied scales, we normalize the data to fall within the range of -1 to 1.

Training procedure follows conventional SL, introduced in \cite{vep_2018}, with slight adjustments to fit the scenario of interest. In more detail, after raw data $\boldsymbol{x}$ is collected, it undergoes pre--processing on the edge device, and an intermediate representation is then sent either directly to the server (via $\mathcal{W}_{\text{ES}}$) or across the backhaul (using a drone, through both $\mathcal{W}_{\text{ED}}$ and $\mathcal{W}_{\text{DS}}$). If the backhaul is utilized, the drone introduces additional processing of the intermediate representation, as depicted in Fig. \ref{fig:sys}. The neural network $F$ is divided into sub-networks across the edge device (comprising one LSTM layer), the drone (comprising two LSTM layers), and the server side (comprising 3 LSTM layers followed by a fully connected (FC) layer) as depicted in Fig. \ref{fig:sys}. The number of LSTM hidden units, denoted as $H$, remains fixed for all conducted experiments, and it is set equal to the length of the intermediate representation $M$, i.e., $H=M=10$. The additional fully connected (FC) layer at the server side also consists of 10 neurons. Training is conducted using a learning rate of $\alpha=0.01$, $\beta_1=0.9$, and $\beta_2=0.999$ on a batch-by-batch basis, with a batch size of 64. We utilize stochastic gradient descent (SGD) with the Adam optimizer, as detailed in Section \ref{model}.

Channel conditions remain fixed during the training phase, following the approach proposed by \cite{OShea_2017}. However, during testing, we assess the model's performance across a range of erasure probabilities $p$. It is important to emphasize that we investigate the impact of different channel conditions introduced during the training phase by varying the training erasure probabilities, $p_{\text{tr}}$, in $\mathcal{W}_{\text{ES}}$ and $\mathcal{W}_{\text{ED}}$, while ensuring that the backhaul between the drone and the server $\mathcal{W}_{\text{DS}}$ always maintains good channel conditions with a small erasure probability (Section \ref{model}). The erasure channel conditions are simulated by incorporating additional dropout layers \cite{hinton_2012} during the training process. These dropout layers replace all three wireless links in Fig. \ref{fig:sys}, akin to the approach outlined in \cite{itahara_2022}, albeit on a symbol basis. We specify a particular dropout probability to regulate the occurrence of channel erasures within our simulations.   

\subsection{Numerical Results}
To examine the behavior of the proposed system under diverse conditions, we consistently assume that one of the channels, either $\mathcal{W}_{\text{ES}}$ or $\mathcal{W}_{\text{ED}}$, introduces significant distortions in the intermediate representation.Additionally, during the testing phase, we set the erasure probability for the more distorted channel to $p_1$, and for the less distorted channel to $p_2$=$p_1-0.3$. Meanwhile, $\mathcal{W}_{\text{DS}}$ remains constant, with its erasure probability set to 0.05 during both the training and testing phases.

\begin{figure}[t]
	\begin{tikzpicture}[spy using outlines=
{rectangle, magnification=4, connect spies}]
  	\begin{semilogyaxis}[width=1\columnwidth, height=7.5cm, 
	legend style={at={(0.35,0.98)}, anchor= north,font=\scriptsize, legend style={nodes={scale=0.9, transform shape}}},
   	legend cell align={left},
	legend columns=1,   	 
   	x tick label style={/pgf/number format/.cd,
   	set thousands separator={},fixed},
   	y tick label style={/pgf/number format/.cd,fixed, precision=2, /tikz/.cd},
   	xlabel={$p_1$},
   	ylabel={MSE},
   	label style={font=\footnotesize},
   	grid=major,   	
   	xmin = 0.3, xmax = 0.9,
   	ymin=0.01, ymax=0.2,
   	line width=0.8pt,
   	tick label style={font=\footnotesize},]

   	\addplot[red, mark=x, mark options={solid}] 
   	table [x={x}, y={y}] {./Figs/d05_s01_b0_mse_edge};
   	\addlegendentry{$f_{\text{E}}+f_{\text{S}}$ -  $\mathcal{W}_{\text{ES}}=0.1$, $\mathcal{W}_{\text{ED}}=0.5$}

    \addplot[dashed, red, mark=x, mark options={solid}] 
   	table [x={x}, y={y}] {./Figs/d05_s01_b0_mse_edge_fc};
   	\addlegendentry{$f_{\text{E}}+\hat{f}_{\text{S}}$ -  $\mathcal{W}_{\text{ES}}=0.1$, $\mathcal{W}_{\text{ED}}=0.5$}

    \addplot[blue, mark=o, mark options={solid}] 
   	table [x={x}, y={y}] {./Figs/d05_s01_b0_mse_server};
   	\addlegendentry{$f_{\text{E}}+f_{\text{D}}+f_{\text{S}}$ -  $\mathcal{W}_{\text{ES}}=0.1$, $\mathcal{W}_{\text{ED}}=0.5$}

    \addplot[dashed, blue, mark=o, mark options={solid}] 
   	table [x={x}, y={y}] {./Figs/d05_s01_b0_mse_server_early};
   	\addlegendentry{$f_{\text{E}}+f_{\text{D}}+\hat{f}_{\text{S}}$ -  $\mathcal{W}_{\text{ES}}=0.1$, $\mathcal{W}_{\text{ED}}=0.5$}
 	\end{semilogyaxis}
	\end{tikzpicture}
	\vspace*{-5mm}
	\caption{MSE versus $p_1$ erasure probabilities: Improved edge-to-server channel conditions ($p_{\text{tr}}$ for $\mathcal{W}_{\text{ED}} > \mathcal{W}_{\text{ES}}$) with $\mathcal{W}_{\text{DS}}=0.05$.}
	\label{Fig_perf_1}
\end{figure}
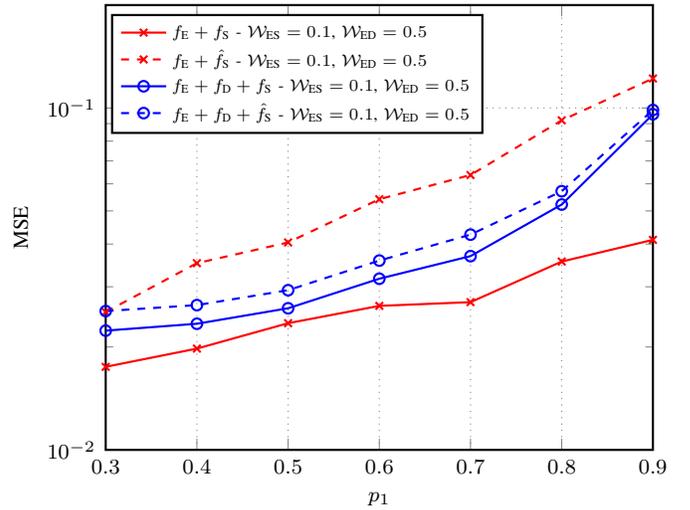

In Fig. \ref{Fig_perf_1}, we compare the MSE performances of the fronthaul and backhaul systems, where the backhaul  ($\mathcal{W}_{\text{ED}}$) significantly alters the intermediate representation due to a high erasure probability. More precisely, training erasure probability for $\mathcal{W}_{\text{ED}}$ is set to 0.5, while for $\mathcal{W}_{\text{ES}}$ to 0.1. Consequently, during the testing phase, erasure probability $p_1$ is associated with $\mathcal{W}_{\text{ED}}$, while the fronthaul link $\mathcal{W}_{\text{ES}}$ is tested with $p_2$. It becomes evident that channel conditions play a significant role in overall system performance. For instance, the fronthaul system, characterized by full server processing under favorable channel conditions (indicated by the red solid line in Fig. \ref{Fig_perf_1}), surprisingly demonstrates superior performance despite its lower processing complexity. It is also notable that, within the backhaul system, additional processing does not lead to the recovery of lost symbols. This is evident from the performances achieved by deploying the full server sub--network (solid blue line in Fig. \ref{Fig_perf_1}), which are similar to those obtained when only the output FC layer is utilized (dashed blue line in Fig. \ref{Fig_perf_1}).

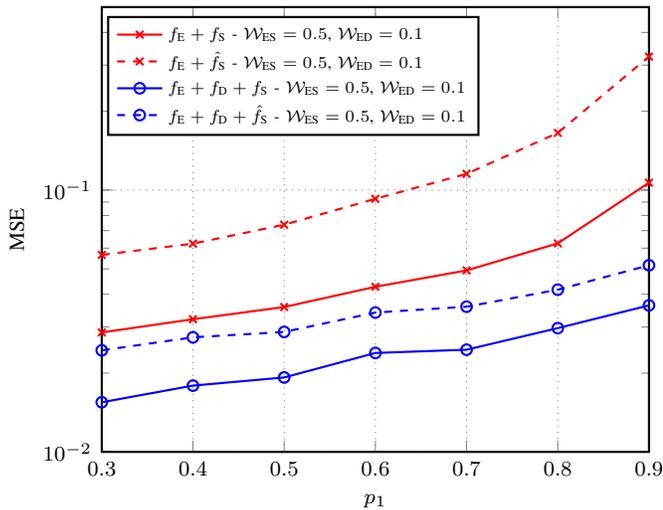
\begin{figure}[t]
	\begin{tikzpicture}[spy using outlines=
{rectangle, magnification=4, connect spies}]
  	\begin{semilogyaxis}[width=1\columnwidth, height=7.5cm, 
	legend style={at={(0.35,0.98)}, anchor= north,font=\scriptsize, legend style={nodes={scale=0.9, transform shape}}},
   	legend cell align={left},
	legend columns=1,   	 
   	x tick label style={/pgf/number format/.cd,
   	set thousands separator={},fixed},
   	y tick label style={/pgf/number format/.cd,fixed, precision=2, /tikz/.cd},
   	xlabel={$p_1$},
   	ylabel={MSE},
   	label style={font=\footnotesize},
   	grid=major,   	
   	xmin = 0.3, xmax = 0.9,
   	ymin=0.01, ymax=0.5,
   	line width=0.8pt,
   	tick label style={font=\footnotesize},]

   	\addplot[red, mark=x, mark options={solid}] 
   	table [x={x}, y={y}] {./Figs/d01_s05_b0_mse_edge};
   	\addlegendentry{$f_{\text{E}}+f_{\text{S}}$ -  $\mathcal{W}_{\text{ES}}=0.5$, $\mathcal{W}_{\text{ED}}=0.1$}

    \addplot[dashed, red, mark=x, mark options={solid}] 
   	table [x={x}, y={y}] {./Figs/d01_s05_b0_mse_edge_early};
   	\addlegendentry{$f_{\text{E}}+\hat{f}_{\text{S}}$ -  $\mathcal{W}_{\text{ES}}=0.5$, $\mathcal{W}_{\text{ED}}=0.1$}

    \addplot[blue, mark=o, mark options={solid}] 
   	table [x={x}, y={y}] {./Figs/d01_s05_b0_mse_server};
   	\addlegendentry{$f_{\text{E}}+f_{\text{D}}+f_{\text{S}}$ -  $\mathcal{W}_{\text{ES}}=0.5$, $\mathcal{W}_{\text{ED}}=0.1$}

    \addplot[dashed, blue, mark=o, mark options={solid}] 
   	table [x={x}, y={y}] {./Figs/d01_s05_b0_mse_server_early};
   	\addlegendentry{$f_{\text{E}}+f_{\text{D}}+\hat{f}_{\text{S}}$ -  $\mathcal{W}_{\text{ES}}=0.5$, $\mathcal{W}_{\text{ED}}=0.1$}
 	\end{semilogyaxis}
	\end{tikzpicture}
	\vspace*{-5mm}
	\caption{MSE versus $p_1$ erasure probabilities: Improved edge-to-drone channel conditions ($p_{\text{tr}}$ for $\mathcal{W}_{\text{ED}} < \mathcal{W}_{\text{ES}}$) with $\mathcal{W}_{\text{DS}}=0.05$.}
	\label{Fig_perf_2}
\end{figure}

Examining Fig. \ref{Fig_perf_2}, we observe that when subjecting our approach to testing conditions resembling real-world scenarios—where the fronthaul link between the edge device and server is highly corrupted, necessitating fallback solutions like the backhaul—we find that additional processing successfully extracts all temporal connections within the time series data. Consequently, the system that integrates all three sub-networks demonstrates superior performance, as indicated by the blue solid line in Fig. \ref{Fig_perf_2}.

The results depicted in Figs. \ref{Fig_perf_1} and \ref{Fig_perf_2} provide compelling evidence that the proposed approach, which integrates backhaul transmission and SL paradigm, offers significant robustness to the server. This robustness enables the server to adapt to unexpected changes in wireless links and define an optimal transmission strategy within various operating conditions. Moreover, it also provides a significant degree of freedom to the server. With this flexibility, the server can incorporate various trade-offs into the implemented strategy, beyond just considering MSE. For instance, it can now factor in parameters such as latency or communication overhead, allowing for a more nuanced and adaptive approach.  

\section{Conclusion}
In this work, we present a novel approach that offers a single, versatile framework for the integration of distributed learning and UAV-assisted relaying in IoT environmental monitoring systems.  The proposed architecture demonstrates significant adaptability to varying channel conditions in IoT systems, offering different trade-offs that can be defined by the server, 
 primarily based on desired performance metrics. Furthermore, the introduction of the SL paradigm optimally balances the computational load between each component (edge device, UAV, and server). In future work, our goal is to incorporate additional parameters into the server decision-making process, such as latency and energy efficiency.

\vspace{12pt}


\begin{thebibliography}{00}

\bibitem{ullo_2020}
S. L. Ullo and G. Sinha, "Advances in smart environment monitoring
 systems using iot and sensors," \emph{Sensors,} vol. 20, no. 11, pp. 3113, May 2020.

\bibitem{mois_2017}
G. Mois, S. Folea, and T. Sanislav, "Analysis of three IoT-based wireless sensors for environmental monitoring," \emph{IEEE Trans. Instrum. Meas.,} vol. 66, no. 8, pp. 20562064, Aug. 2017.

\bibitem{mozafir_2019}
M. Mozaffari, W. Saad, M. Bennis, Y.-H. Nam, and M. Debbah,
 “A tutorial on UAVs for wireless networks: Applications, challenges,
 and open problems,” \emph{IEEE Commun. Surveys Tuts.,} vol. 21, no. 3,
 pp. 2334–2360, 3rd Quart., 2019.

\bibitem{sabze_2022}
 J. Sabzehali, V. K. Shah, Q. Fan, B. Choudhury, L. Liu, and
 J. H. Reed, “Optimizing number, placement, and backhaul connectiv
ity of multi-UAV networks,” \emph{IEEE Internet Things J.,} vol. 9, no. 21,
 pp. 21548–21560, Nov. 2022.

\bibitem{zheng_2016}
Y. Zeng, R. Zhang, and T. J. Lim, “Wireless communications with
 unmanned aerial vehicles: opportunities and challenges,” \emph{IEEE Commun. Mag.,} vol. 54, no. 5, pp. 36–42, May 2016.

\bibitem{galkin_2018}
B. Galkin, J. Kibilda, and L. A. DaSilva, “Backhaul for low-altitude
 UAVs in urban environments,” in \emph{Proc. IEEE Int. Conf. Commun. (ICC),} 2018, pp. 1–6.

\bibitem{yaacoub_2021}
E. Yaacoub and M.-S. Alouini, "Efficient fronthaul and backhaul connectivity for IoT traffic in rural areas", \emph{IEEE Internet Things Mag.,} vol. 4, no. 1, pp. 60-66, Mar. 2021.

\bibitem{zhang_2019}
L. Zhang and N. Ansari, "Optimizing the deployment and throughput
 of dbss for uplink communications," \emph{IEEE Open J. Veh. Tech.,} vol. 1, pp. 18–28, 2019.


\bibitem{fouda_2018}
 A. Fouda, A. S. Ibrahim, I. Guvenc, and M. Ghosh, "UAV-based
 in-band integrated access and backhaul for 5G communications," in
 \emph{Proc. IEEE Conf. Veh. Tech.,} 2018, pp. 1–5.

\bibitem{selim_2018}
M. Y. Selim and A. E. Kamal, "Post-disaster 4G/5G network rehabil
itation using drones: Solving battery and backhaul issues," in \emph{Proc.
 IEEE Globecom Workshops (GC Wkshps),} 2018, pp. 1–6.

\bibitem{gupta_2018}
O. Gupta and R. Raskar, "Distributed learning of deep neural network over multiple agents," \emph{J. Netw. Comput. Appl.,} vol. 116, pp. 1--8, Aug. 2018.

\bibitem{vep_2018}
P. Vepakomma, O. Gupta, T. Swedish, and R. Raskar, "Split learning for health: Distributed deep learning without sharing raw patient data," 2018, arXiv:1812.00564. [Online]. Avaliable: https://doi.org/10.48550/arXiv.1812.00564

%\bibitem{poirot_2019}
%M. G. Poirot, P. Vepakomma, K. Chang, J. Kalpathy-Cramer,
% R. Gupta, and R. Raskar,"Split learning for collaborative deep learning in healthcare," 2019, arXiv:1912.12115.[Online]. Available:
% https://doi.org/10.48550/arXiv.1912.12115

 \bibitem{husken_2002}
 M. Hüsken and P. Stagge, "Recurrent neural networks for time series classification," \emph{Neurocomputing}, vol. 50, pp. 223--235, Jan. 2003.

\bibitem{b13}
 S. Hochreiter and J. Schmidhuber, ``Long Short-Term Memory,'' \emph{Neural Computation,} vol. 9, pp. 1735–1780, 1997.

 \bibitem{abua_2020}
 S. Abuadbba, K. Kim, M. Kim, C. Thapa, S. A. Camtepe, Y. Gao, H. Kim, and S. Nepal, "Can we use split learning on 1d cnn models
 for privacy preserving training?" in \emph{Proc. 15th ACM Asia, ser. ASIA CCS
 ’20.} New York, NY, USA: Association for Computing Machinery,
 pp. 305--318, 2020.

  \bibitem{Gao_2020}
 Y. Gao, M. Kim, S. Abuadbba, Y. Kim, C. Thapa, K. Kim, S. A. Camtep, H. Kim, and S. Nepal, "End-to-end evaluation of federated learning and split learning for internet of things," in \emph{Proc. IEEE 2020 Int. Symp. on Rel. Distrib. Syst.,} Shanghai, China, Sep. 21-24,
 pp. 91--100., 2020.

 \bibitem{zhang_2024}
 W. Zhang, T. Zhou, Q. Lu, Y. Yuan, A. Tolba, and W. Said, "FedSL: A Communication Efficient Federated Learning With Split Layer Aggregation," \emph{IEEE Internet Things J.}, Early Access

 \bibitem{koda_2020}
 Y. Koda, J. Park, M. Bennis, K. Yamamoto, T. Nishio, M. Morikura,
 and K. Nakashima, "Communication-efficient multimodal split learning
 for mmWave received power prediction," \emph{IEEE Commun. Lett.}, vol. 24,
 no. 6, pp. 1284--1288, June 2020.

 \bibitem{jiang_2022}
 L. Jiang, Y. Wang, W. Zheng, C. Jin, Z. Li, and G. S. Teo, "LSTMSPLIT: effective SPLIT learning based LSTM on sequential time-series data," 2022, arXiv: cs.LG/2203.04305. [Online]. Available: 
https://doi.org/10.48550/arXiv.2203.04305

\bibitem{abedi_2023}
 A. Abedi and S. S. Khan, "Fedsl: Federated split learning on distributed sequential data in recurrent neural networks," \emph{Multimed. Tools. Appl.}, vol. 83, pp. 28891--28911, Sept. 2023.

\bibitem{adam}
  D. P. Kingma and J. L. Ba, "Adam: A method for stochastic optimization," in \emph{Proc. Int. Conf. on Learn. Representation,} May 7--9, pp. 1-41, 2015.

  \bibitem{itahara_2022}
 S. Itahara, T. Nishio, Y. Koda, and K. Yamamoto, "Communication-Oriented Model Fine-Tuning for
 Packet-Loss Resilient Distributed Inference
 Under Highly Lossy IoT Networkss," \emph{IEEE Access}, vol.10, pp. 14969--14979, 2022.

 \bibitem{OShea_2017} %postoji
T. O’Shea and J. Hoydis, "An introduction to deep learning for the physical layer," \emph{IEEE Trans. Cogn. Commun. Netw.,} vol. 3, no. 4, pp. 563-575, Dec. 2017.

\bibitem{hinton_2012}
G. E. Hinton, N. Srivastava, A. Krizhevsky, I. Sutskever, and R.R.Salakhutdinov, "Improving neural networks by preventing co-adaptation of feature detectors," 2012,	arXiv: cs.NE/arXiv:1207.0580. [Online]. Available: https://arxiv.org/abs/1207.0580

\end{thebibliography}
\end{document}